\documentclass[prl,showpacs,twocolumn]{revtex4}

\usepackage{amsmath}
\usepackage{epsfig}          

\begin{document}
\title{Phase Coexistence in Driven One Dimensional Transport}
\author{A. Parmeggiani (1), T. Franosch (1), and E. Frey (1,2)}
\affiliation{(1) Hahn-Meitner Institut, Abteilung Theorie, Glienicker
  Str. 100, D-14109 Berlin, Germany\\ (2) Fachbereich Physik, Freie
  Universit\"at Berlin, Arnimallee 14, D-14195 Berlin, Germany}
\date{\today} 
\begin{abstract}
  We study a one-dimensional totally asymmetric exclusion process with
  random particle attachments and detachments in the bulk. The
  resulting dynamics leads to unexpected stationary regimes for large
  but finite systems. Such regimes are characterized by a phase
  coexistence of low and high density regions separated by domain
  walls. We use a mean-field approach to interpret the numerical
  results obtained by Monte-Carlo simulations and we predict the phase
  diagram of this non-conserved dynamics in the thermodynamic limit.
\end{abstract}
\maketitle

Even some of the simplest driven diffusive systems in one dimension
show surprisingly rich and complex behavior which is rather unexpected
when looked at with experience gained from equilibrium
phenomena~\cite{schmittmannbook}.  A particularly illuminating example
are boundary-induced phase transitions in driven one-dimensional (1D)
transport processes, such as the Totally Asymmetric Simple Exclusion
Process (TASEP). The model, originally proposed in~\cite{mcdo68},
consists of particles hopping unidirectionally with hard-core
exclusion along a 1D lattice. Due to conservation of the particle
current in the bulk, the rates of incoming or outgoing particles at
the boundaries drive the system to non-trivial stationary
states~\cite{krug91}.  The resulting phase diagram shows continuous
and discontinuous transitions of the average density of particles in
the limit of large system sizes. These results were obtained first in
mean-field theory and then extended when a complete analytical
solution was presented solving explicitely the recursion relations of
the model or using a matrix product ansatz technique~\cite{solutions}.
 
The TASEP is one out of many examples for driven systems with
stationary non-equilibrium states, which cannot be described in terms
of Boltzmann weights. This has to be contrasted with processes like
the bulk adsorption/desorption kinetics of particles on a lattice
coupled to a reservoir (``Langmuir Kinetics'', LK), whose stationary
state is well described within standard concepts of equilibrium
statistical mechanics.  Here, particles adsorb at an empty site or
desorb from an occupied one with fixed respective kinetic rates
obeying detailed balance. The bulk density profile at equilibrium is
described by a {\it Langmuir isotherm}, determined solely by the ratio
of the two kinetic rates~\cite{fowlerbook}, as given by the Gibbs
ensemble. Due to the presence of the particle reservoir there is no
conservation of particles and no net particle current in the bulk. It
is interesting to ask what can be expected in coupling two processes
which have genuinely different dynamics and stationary states, like
TASEP with open boundaries and LK.

In this letter, we relax the constraint that the conserved dynamics in
the bulk imposes to the TASEP by allowing particle attachment and
detachment. We are interested in the limit where the kinetic rates are
such that the incoming and outgoing fluxes of particles at the boundaries
and in the bulk are comparable.  This implies that a particle, injected at
the boundary or attached somewhere in the bulk, remains long enough on
the lattice to move a finite fraction of the total system size. New
phenomena are expected in the regime of competition between TASEP and
LK for a large but finite system. Of course, the dynamics in an
infinitely large system would be completely dominated by the bulk
adsorption and desorption rates. It turns out that the presence of the kinetic
rates significantly change the picture of TASEP, producing a
completely reorganized phase diagram. We shall show by computer
simulations and mean-field arguments that, in this non-conserved
dynamics, one can have phase coexistence where low and high density
phases are separated by stable discontinuities in the density profile.

The model we discuss here is directly inspired by the unidirectional
motion of many motor proteins along cytoskeletal
filaments~\cite{albertsbook}. Motors advance along the filament while
attachment and detachment of motors between the cytoplasm and the
filament occur~\cite{note0}. Recently, it has been shown that such
dynamics can be relevant for modeling the filopod growth in eukaryotic
cells produced by motor proteins interacting within actin
filaments~\cite{krus02}.
\begin{figure}[ht!]
\vspace{-0.1cm}
\epsfig{file=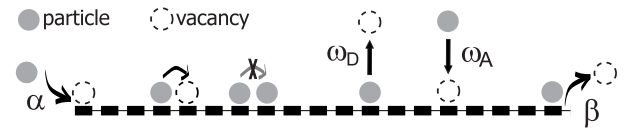,width=8. cm}
\vspace{-0.2cm}
\caption{TASEP scheme with bulk attachment/detachment.}
\label{f:model}
\vspace{-0.2cm}
\end{figure}

We consider a 1D lattice composed of sites $i\!=\!1,...,N$
(Fig.~\ref{f:model}).  The configurations are described in terms of
occupation numbers $n_i\! =\! 1$ for a site occupied by a particle and
$n_i\! =\! 0$ for an empty site (vacancy).  The dynamics is determined
by a master equation for the probabilities to find a particular
configuration $\{n_i\}$. We apply the following dynamical rules. For
each time step, a site $i$ is chosen at random. A particle at site $i$
can jump to site $i+1$ if unoccupied (we fix units of time by putting
this rate equal to unity). In the bulk $i\!=\!2,..,N-1$, a particle
can also leave the lattice with site-independent detachment rate
$\omega_D$ or fill the site (if empty) with a rate $\omega_A$ by
attachment. At the boundaries, a particle can fill a vacancy with a
rate $\alpha$ at site $i\!=\!1$, or a vacancy can be formed by
removing a particle from the lattice with a rate $\beta$ at site
$i\!=\!N$. We refrain from giving explicitly the master equation for
the probabilities. Correlations induced into the many-particle problem
can be conveniently studied within an operator representation in Fock
space~\cite{schutzreview}.  Then the equations of the bulk dynamics
read:
\begin{subequations}\label{eq:micro}
\begin{equation}
\label{eq:Dnew}
  \frac{d n_i}{dt} = 
  n_{i-1}(1-n_{i}) - n_i(1-n_{i+1}) + 
  \omega_A(1-n_i) - \omega_D n_i ,
\end{equation}
while at the boundaries one obtains:
\begin{equation}
\label{eq:boundaries}
\begin{array}{l}
\displaystyle{{d n_1}/{dt} = \alpha(1-n_{1}) - n_1(1-n_{2})} \, ,\\
\displaystyle{{d n_N}/{dt} = n_{N-1} (1-n_{N}) - \beta n_N}  \, .
\end{array}
\end{equation} 
\end{subequations}
By taking averages \cite{note1} one observes that in order to compute the time
evolution of $\langle n_i(t) \rangle$ one needs the corresponding
averages of higher order correlations. In order to obtain an exact
solution, elaborate techniques are necessary. We restrict the
discussion to Monte-Carlo simulations (MCS) and a
mean-field approximation (MFA) which we shall apply below.
 
The system exhibits a particle-hole symmetry in the following sense.
A jump of a particle to the right corresponds to a vacancy move by one
step to the left.  Similarly, a particle entering the system at the
left boundary can be interpreted as a vacancy leaving the lattice, and
vice versa for the right boundary. Attachment and detachment of
particles in the bulk is mapped to detachment and attachment of vacancies,
respectively.

We are interested in large system sizes ($N \gg 1$) and, eventually,
in the ``thermodynamic limit'' $N \to \infty$. In this case, the study
of the competition between bulk and boundary dynamics needs that
the kinetic rates decrease simultaneously with the system
size.  More precisely, we define the ``reduced'' rates $\Omega_A$ and
$\Omega_D$ as $\Omega_{A}\!=\!\omega_{A} N$ and $\Omega_D\!=\!\omega_D N$,
keeping $\Omega_A, \Omega_D, \alpha, \beta$ fixed as $N\!\to\!\infty$.
Note that the binding constant $K\!\!=\!\omega_A/\omega_D$ remains
unchanged when passing to the thermodynamic limit. Moreover,
for $\omega_A\!=\!\omega_D\!=\!0$, one arrives back at the TASEP
respecting the same particle-hole symmetry described above.
\begin{figure}
\epsfig{file=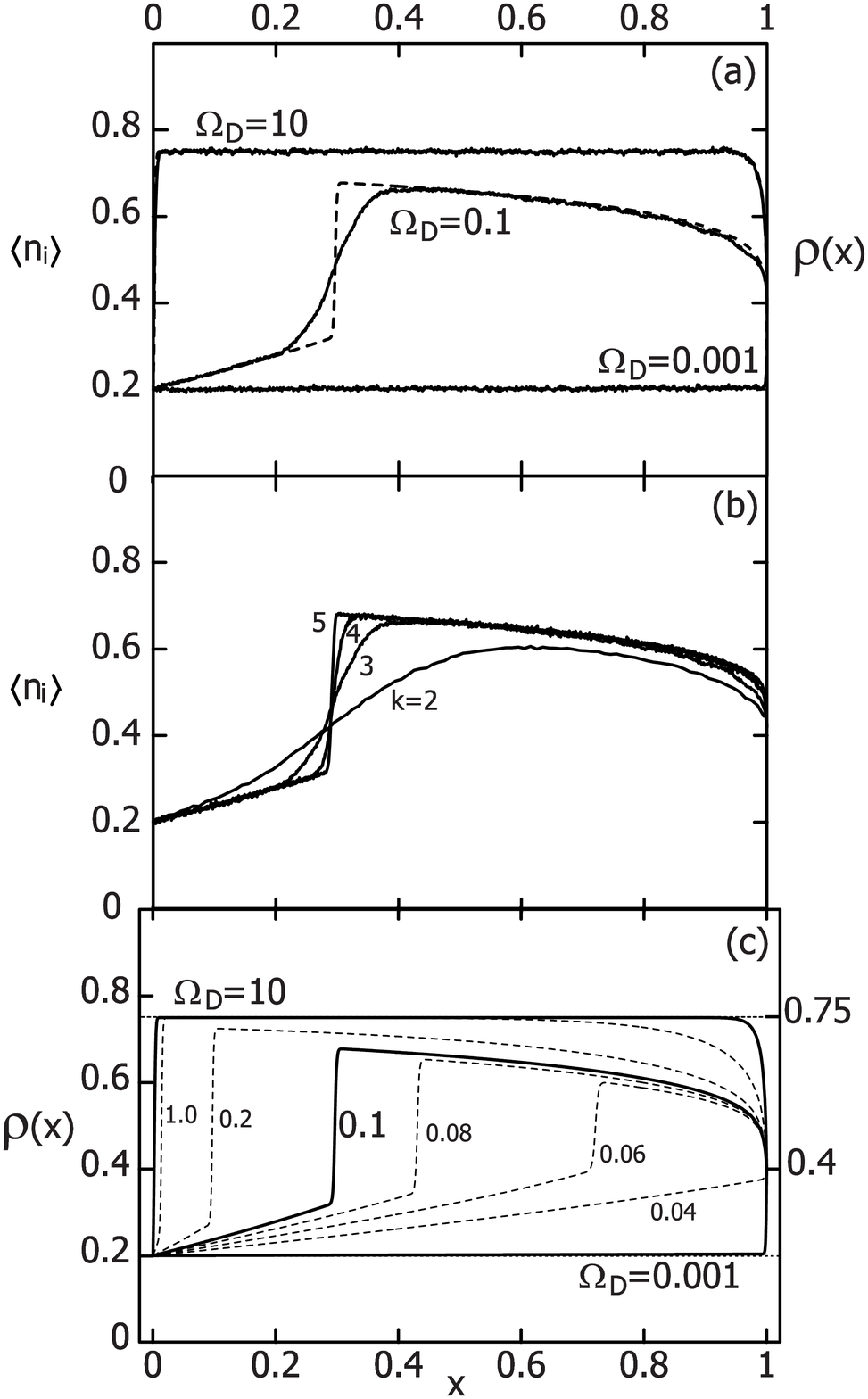,width=9. cm}
\caption{{\bf (a)} Average density profile $\langle n_i \rangle$ computed by
MCS (continuous line) and average density profile $\rho(x)$ computed
by numerical integration of MFA stationary state equations
(\ref{eq:mfa}) (dashed line) in the rescaled variable $x\!=\!i/N$ for
$N \! = \! 10^3$ with $\alpha \!= \!0.2,\,\beta\!=\!0.6,\, K\!=\!3$
and different kinetics rates $\Omega_D$ indicated in the graph.{\bf(b)}
MCS average density profile for different system sizes, 
same $\alpha, \beta, K$ as before, and $\Omega_D\!=\!0.1$. The
width of the steep rise decreases with increasing system sizes
$N\!=\!10^k$ with $k\!=\!2,3,4,5$ indicated in the graph. {\bf(c)} MFA average
density profile for $\varepsilon\!=\!10^{-3}$,
same $\alpha, \beta, K$ as before, and different kinetic rates
$\Omega_D$ indicated in the graph.  The horizontal dashed line for
$\rho(x)\!=\!0.75$ represents the Langmuir isotherm for $K\!=\!3$.}
\label{f:densities}
\vspace{-0.2cm}
\end{figure}

We have performed extensive computer simulations~\cite{note2} to
obtain the average density profile in the stationary state. We
illustrate typical phenomena by following a path in parameter space
along curves with fixed $\alpha, \beta$, and $K$ while increasing
$\Omega_D \!= \!\Omega_A / K$.  Fig.~\ref{f:densities}(a) shows the density
profile for three different values of the kinetic rates.  At small
kinetic rates, $\Omega_A,\Omega_D \ll \alpha,\beta$, the average
density $\langle n_i \rangle$ in the bulk is practically constant and
close to the low-density value predicted by the TASEP, $\langle n_i
\rangle \!=\! \alpha$. Conversely, at high kinetic rates the bulk profile
is structureless and essentially determined by the well-known ratio
$K/(1+K)$ of Langmuir equilibrium density~\cite{note3}. A new feature
appears for intermediate rates, $\Omega_D \!=\! 0.1$, precisely when bulk
and boundary dynamics compete. The density exhibits a non-monotonic
structure in bulk, characterized by a region of low and high density
connected by a steep rise.

Fig. \ref{f:densities}(b) shows the density profile for different system
sizes. One observes a decrease of the width of the transition region
as the number of sites is increased.  The simulation suggests a
discontinuity of the profile in terms of the rescaled variable $x \!=\!
i/N$ upon approaching the infinite system limit. A preliminary finite
size scaling analysis is compatible with a rescaled transition width
which scales as $N^{-\nu}$ with $\nu\simeq 0.5 .$ This is very
different from the mean-field result, $\nu_{\rm MFA} \!=\!1$~\cite{note4}.
Thus we have identified an intermediate parameter range where low and
high density phases coexist separated by a sharp domain wall (DW). We
also find that the discontinuity in the density seems to be stable or
at least localized in a small region compared to the system
size~\cite{note5}. This has to be contrasted with the domain wall
(``shock'') found in the TASEP right at the phase boundary between the
high and low density phase ($\alpha\!=\!\beta \!< \!1/2$) which is
de-localized and moves as a random walker once it is far from the
system boundaries~\cite{schutzreview}.

Phenomena like phase separation/coexistence have previously been
observed in non-homogeneous systems with open boundaries like TASEP
with isolated localized defects~\cite{lebo92,kolo}. The location of
the domain walls are expected and found to be identical to the defect
positions. In contrast, the location of the DW in our homogenous model
is self-tuned and determined by the values of the kinetic rates (see
below). In systems with periodic boundary conditions (which are not
the subject of this letter) phase separation has been found in TASEP
with a blockage \cite{schu93}, quenched  disorder
\cite{barm98}, or in homogeneous systems with multi-species particle
dynamics \cite{evan98} (see for a general criterion \cite{kafr02}).

To rationalize all these findings we have developed a mean-field
theory. Defining $\rho_i \!=\! \langle n_i \rangle$, the MFA consists of
taking the average of Eqs.~(\ref{eq:Dnew},\ref{eq:boundaries}) and
factorizing the two-site correlations, $\langle n_i n_{i+1}\rangle \!=\!
\rho_i \rho_{i+1}$.  Then Eqs.
(\ref{eq:Dnew},\ref{eq:boundaries}) display the same form provided
that the binary occupation number $n_i$ is replaced by the continuous
variable $\rho_i$ with $0 \!\leq\! \rho_i \!\leq\! 1$. The equations are now
interpreted as ordinary differential equations.

To obtain an analytically tractable system of equations we have
coarse-grained the discrete lattice with lattice constant $\varepsilon\!
=\! L/N$ to a continuum. For fixed total length $L\!=\!1$ and $N\!\to\!
\infty, \varepsilon\!\to\!0$ one gets the nonlinear differential
equation for the average profile in the stationary state,
\begin{eqnarray}
\label{eq:mfa}
 \frac{\varepsilon}{2}\partial_x^2 \rho 
 +  (2 \rho -1)\partial_x \rho +  \Omega_A (1-\rho) -  \Omega_D \rho = 0 \, ,
\end{eqnarray}
where positions are measured by the rescaled variable $x\! =\! i/N, \,
\, 0 \!\leq x\! \leq\! 1$.  Equations (\ref{eq:boundaries}) translate
now to boundary conditions for the density field, $\rho(0)\!=\!\alpha$
and $\rho(1)\!=\! 1-\beta.$ One observes that MFA respects the
particle-hole symmetry mentioned above, provided that when $\rho(x)
\mapsto 1 - \rho(1-x)$ one interchanges $\alpha \!\leftrightarrow
\!\beta, \, \,\Omega_A \!\leftrightarrow \!\Omega_D$.  Due to this
property we can restrict the discussion to the case $\Omega_A \!>
\!\Omega_D$ \cite{note6}.  The numerical mean-field solutions are
included in Fig. \ref{f:densities}(a) for different values of
$\Omega_D$. We find good agreement of MFA compared with MCS for the
full range of kinetic rates in the limit of large $N$.

In analogy with fluid dynamics, to describe these results one
considers an effective current density which for our problem reads $j
\!=\! - (\varepsilon/2) \partial_x \rho + \rho (1 - \rho)$. Abbreviating
the fluxes from and to the reservoir by ${\cal F}_A\!=\!\Omega_A
(1-\rho)$ and ${\cal F}_D\!=\!\Omega_D \rho$, Eq. (\ref{eq:mfa}) can be
read as a balance equation: $\partial_x j\!=\!{\cal F}_A - {\cal
F}_D$. Since there are two boundary conditions one has to be careful
when discarding the second derivative in (\ref{eq:mfa}) for a 
small prefactor $\varepsilon$. The average profile is then governed
by similar physics as the Burgers' equation in the inviscid limit
\cite{lebo92}.  Generically one expects shocks (here, ``domain
walls'', DW) in the bulk and density layers at the boundaries
(``boundary layers''). Crossing a DW, the current $j$ remains
continuous in the limit $\varepsilon\!\to\!0$, while boundary layers form
whenever the density associate to the bulk current does not fit the
boundary condition.  To better understand these features, we have explored
the dependence of the density profile $\rho (x)$ on $\Omega_D$ for
fixed $\alpha, \beta$ and $K$ (see Fig.  \ref{f:densities}(c)).  For small
kinetic rates, $\Omega_D\!=\!10^{-3}$, the profile is close to the one
expected from TASEP, with a boundary layer bridging the bulk density
up to $\rho\!=\!1-\beta$ (not resolved in Fig. \ref{f:densities}(c)).
Increasing $\Omega_D$ the slope of the bulk density increases. For
$\Omega_D\!>\!0.05$, MFA connects a region of low density (LD), i.e.
$\rho(x)\!<\!1/2$ to a high density region (HD), $\rho(x)\!>\!1/2$, by a
DW. Whereas the solution close to the left boundary is smooth, one
finds a boundary layer at the right end bridging densities $\rho\!=\!1/2$ 
down to $\rho\!=\!1-\beta$.  For larger $\Omega_D$ the DW moves to
the left, while the slope of the LD region increases and the HD
profile flattens approaching the Langmuir density value $K/(K+1)$.
For $\Omega_D\!>\!1$ the DW remains practically localized at the left
boundary. Note that the DW location strongly depends on typical values
of the bulk kinetic rates when they are comparable with the boundary
rates $\alpha$ and $\beta$.

In the inviscid limit $\varepsilon\!\to\!0$ the complete phase
diagram can be obtained analytically within MFA, up to some
treatment of the density discontinuities. Interestingly, the solution
found is never given by either a constant low/high density profiles as
in TASEP or the Langmuir isotherm, but by a completely new set of
solutions \cite{parm02}.

\begin{figure}
\vspace{-0.1cm}
\epsfig{file=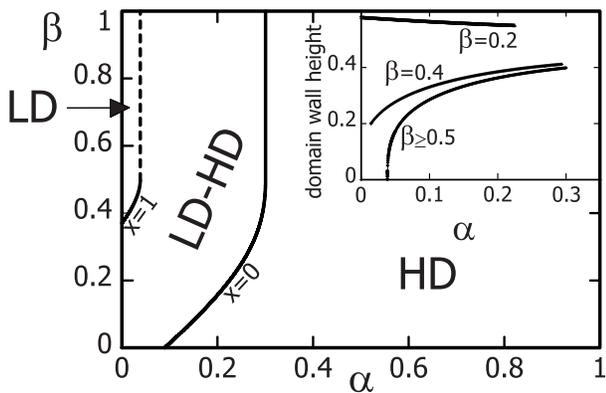,width=8. cm}
\vspace{-0.2cm}
\caption{Phase diagrams obtained by the exact solution of the
  stationary mean-field equation (\ref{eq:mfa}) in the inviscid limit
  for $K\!=\!3$ and $\Omega_D\!=\!0.1$.  The inset shows the dependence of
  the DW amplitude on $\alpha$ for different values of
  $\beta$.}
\vspace{-0.2cm}
\label{f:phase}
\end{figure}

The mean-field analytical solution allows to draw the phase diagram
and compare it to TASEP. Fig. \ref{f:phase} represents a cut through
the phase diagram for $\Omega_D\!=\!0.1$ and $K\!=\!3$ with $\alpha$ and $\beta$
used as control parameters.
One finds an extended LD-HD coexistence region separating a LD and a
HD phase. At the boundaries of the coexistence region, the DW
between the low and high density phases are located in the proximity
of the open ends of the 1D lattice. For small $\alpha$ the DW
develops at the right end, $x\!=\!1$ and moves to the left as $\alpha$
increases. At the phase boundary between the coexistence region and
the HD phase, the DW is located at the left end of the
lattice, $x\!=\!0$. In both cases, when the DW enters and leaves
the lattice, it matches with a boundary layer connecting the bulk
density with the density given by the corresponding boundary
condition. (In this case,  boundary layers appear only for
$\alpha\!>\!1/2$ and/or $\beta\!>\!1/2$).   MFA shows that the bulk
density profile becomes independent of $\beta$ for $\beta\!>\!1/2$.
Hence, in this regime for a given $\alpha$ only the magnitude of the
boundary layers changes, but not the profile of the bulk density.
This explains the vertical phase boundaries of the coexistence region
for $\beta\!\ge\!1/2$.

In addition to its location the DW is also characterized by its height
(see Fig.~\ref{f:phase}). For $\beta\!<\!1/2$ we find that the height
discontinuously jumps to a finite value upon entering the coexistence
region from the LD phase. This has to be contrasted with the case
$\beta\!\ge\!1/2$ where the phase transition from the LD to the
coexistence phase is characterized by a continuous increase in the
height of the DW (compare the dashed line in Fig.~\ref{f:phase}). In
MFA we find that both DW amplitude and position exhibit power law
behavior with $(\alpha-\alpha_c)^{1/3}$ and $(\alpha-\alpha_c)^{2/3}$,
respectively.  At the phase boundary to the HD phase the DW height
always jumps to zero discontinuously.

Working out the complete rich scenario for different $\Omega_A$ and
$\Omega_D$ in the limit $N\!\gg\!1$ needs a detailed analysis of the
phase diagram~\cite{parm02}. We just mention that the original maximal
current phase of the TASEP appears for $\Omega_A\!=\!\Omega_D$ only, where
the Langmuir density is valued to $1/2$. This can be proved
numerically as well as analytically.  Conversely, as soon as $\Omega_A\!\ne\!\Omega_D$, 
MFA predicts that such a phase continuously disappears in
favor of a HD phase if $\Omega_A\!>\!\Omega_D$ (respectively LD phase if
$\Omega_A\!<\!\Omega_D$).

The authors thank P. Benetatos and J.E. Santos for useful
discussions and comments. This work was partially funded by the DFG
under contract nos. FR 850/4-1 and SFB 413. A.P. is supported by a Marie-Curie
Fellowship no. HPMF-CT-2002-01529.


\begin{thebibliography}{99}
\bibitem{note1} The notion of averaging is non-trival since the system
  is far from equilibrium. Nevertheless, one expects the initial state
  to relax to a stationary ensemble over which averages
  should be taken.
\bibitem{note2} MCS were performed with random sequential updating in
discrete time: at each time step, a site is chosen at random and the
dynamical rules are applied. Time and sample averages have been
evaluated.  The resulting profiles coincide in both averaging
procedures for given parameters and different system sizes. In
simulations for Figs.~\ref{f:densities}(a) and \ref{f:densities}(b)
stationary profiles have been obtained over $10^5$ time averages (with
a typical time interval of $\sim 10~N$ between each average step).
\bibitem{note3} In this case, during the detachment and attachment
  processes, we consider a constant particle concentration in the
  reservoir $c_r$ (i.e. $c_r\!=\!1$).
\bibitem{note4} Upon introducing $x^\prime \!=\! x/\varepsilon $, a
scaling analysis of Eq. (\ref{eq:mfa}) shows that the domain wall width scales as 
$\nu_{\rm MFA}\!=\!1$ \cite{parm02}.
\bibitem{note5} As a simple test, we artificially shifted the front of
  the discontinuity and let the system evolve in time. The
  discontinuity always moved back to its initial position.
\end{thebibliography}
\end{document}